\documentclass[12pt]{article}

\usepackage{amsmath}
\usepackage{graphicx}

\title{ Dynamics of quark-gluon plasma from Field correlators }
\author{A. Di Giacomo$^{a)}$, E. Meggiolaro$^{a)}$, Yu.A. Simonov$^{b)}$,
 A.I. Veselov$^{b)}$\\
$^{a)}$ Dipartimento di Fisica ``E. Fermi'' and INFN, Sezione di Pisa,\\
Largo Pontecorvo 3, I--56127 Pisa, Italy\\
$^{b)}$ State Research Center\\Institute of Theoretical and
Experimental Physics, \\ Moscow, 117218 Russia}
\date{}

\newcommand{\beq}{\begin{eqnarray}}
\newcommand{\eeq}{\end{eqnarray}}
\newcommand{\be}{\begin{equation}}
\newcommand{\ee}{\end{equation}}

\def\la{\mathrel{\mathpalette\fun <}}

\def\fun#1#2{\lower3.6pt\vbox{\baselineskip0pt\lineskip.9pt
\ialign{$\mathsurround=0pt#1\hfil ##\hfil$\crcr#2\crcr\sim\crcr}}}

\newcommand{\ver}{\mbox{\boldmath${\rm r}$}}

\newcommand{\vep}{\mbox{\boldmath${\rm p}$}}

\newcommand{\veS}{\mbox{\boldmath${\rm S}$}}
\newcommand{\veL}{\mbox{\boldmath${\rm L}$}}

\newcommand{\lan}{\langle}
\newcommand{\ran}{\rangle}
\begin{document}

\maketitle

\begin{abstract}

It is argued that strong dynamics in the quark-gluon plasma and
bound states of quarks and gluons is mostly due to nonperturbative
effects described by field correlators.
The emphasis in the paper is made on two explicit calculations of
these effects from the first principles -- one analytic using
gluelump Green's functions and another using independent lattice
data on correlators. The resulting hadron spectra are
investigated in the range $T_c\leq T < 2 T_c$.
The spectra of charmonia, bottomonia, light $s \bar s$ mesons,
glueballs and  quark-gluon states calculated numerically are in
general agreement with lattice MEM data. The possible role of these
bound states in the thermodynamics of quark-gluon plasma is
discussed.
\end{abstract}

\section{Introduction}

The importance of nonperturbative dynamics in QCD, which is
illustrated by the phenomena of confinement and chiral symmetry
breaking, was also proposed some time ago for the deconfined phase
\cite{1} -- \cite{3}. In the framework of the Vacuum Correlator
Method (VCM) (sometimes also called Stochastic Vacuum Model)
\cite{4} it was shown (see \cite{5} for a review), that four
dominant correlators of the QCD vacuum $D^{(E,H)}(x)$,
$D_1^{(E,H)}(x)$ define with few percent accuracy all dynamics of
the QCD vacuum and hadrons both below and above $T_c$, and
$D^{(E)}(x)$ is responsible for confinement, with the string
tension $\sigma^{(E)} = 1/2\int D^{(E)}(x) d^2 x$. Above $T_c$ as
was predicted in \cite{1} -- \cite{3}, $D^E(x)$ vanishes, while
$D_1^{(E)}(x)$ stays nonzero and may support the nonperturbative
dynamics at $T \geq T_c$ (together with magnetic corrections due
to $D^{(H)}(x), D_1^{(H)}(x)$).

At approximately the same time, starting from 1992, a careful
study of field correlators was performed by the Pisa group
\cite{6,7} resulting in the explicit forms of four independent
field correlators $D^{(E)}, D_1^{(E)}, D^{(H)}$ and $D_1^{(H)}$
both below and above $T_c$. It was concluded from these results
that indeed $D^{(E)}$ vanishes at $T \geq T_c$, while three other
correlators stay nonzero at least till $T = 1.26 T_c$.

It was suggested in \cite{1} that $D_1^{(E)}(x)$ is responsible
for possible bound states of quarks and gluons in the quark-gluon
plasma, which was then called ``the strong interacting quark-gluon
plasma''. Recently this phenomenon was observed in the lattice at
$T \geq T_c$ in the form of bound states of light $q \bar q$
mesons \cite{8}, heavy quarkonia $(c\bar c)$ \cite{9} -- \cite{13}
and three-quark stetes \cite{14}. Moreover, the thermodynamic
quantities associated with the $Q\bar Q$ system, namely free
energy $F^{(Q\bar Q)}(r, T)$ and internal energy $U^{(Q \bar
Q)}(r,T)$ have been measured \cite{13, 15, 16} for the $Q \bar Q$
distance $r$ in the interval $0 \leq r \la 2$ fm, showing large
asymptotic value, e.g. $F(\infty, T_c) \approx 600$ MeV for $n_f
=2$ \cite{16}.

These latter quantities can be explained only by nonperturbative
effects, since perturbative OGE potential, even with increased
$\alpha_S(r)$, cannot produce similar effect.

At the same time, it is characteristic for the static potential
$V_1(r,T)$ produced by the correlator $D_1^E(x)$, that it indeed
gives rise to the constant term $V_1(\infty, T)$ in the $Q \bar Q$
interaction at large distances, which can be viewed upon as the
sum of constant selfenergies of $Q$ and $\bar Q$.

It was argued in the recent paper \cite{17} by one of the authors
(Yu.S.), that this strong interaction observed in the quark-gluon
plasma on the lattice in \cite{8} -- \cite{16}, and possibly seen
in the ion-ion collisions at RHIC \cite{18}, can be explained by
the correlator $D_1^E(x)$. To check this prediction in \cite{17}
the analytic form of $D_1^E(x)$ was used calculated before in
\cite{19}, exploiting the connection of field correlators to the
gluelump Green's functions. In this way the static $Q \bar Q$
potential $V_1(r,T)$ was computed from $D_1^E(x)$ and compared to
the lattice free energy $F^{(Q\bar Q)}(r,T)$. The resulting
agreement has demonstrated that the $D_1^{(E)}(x)$ can be
considered as a calculable source of strong interaction above
$T_c$.

Based on $V_1(r,T)$ the possibility of bound states of heavy
quarkonia, glueballs and baryons was established, using the
Bargmann condition for the static potential.

It is a purpose of the present paper to calculate explicitly the
possible bound states of quarks and gluons using directly the
previously calculated $D_1^{(E)}(x)$ on the lattice as an input.
In doing so a new fit of lattice data is done, using the
analytically calculated form of $D_1^{(E)}(x)$, found in
\cite{6,7}. For comparison we also use static potential obtained
from the analytic expression of $D_1^{(E)}$, found in \cite{19} for $T
\leq T_c$ and analytically continued to $T \geq T_c$.

With the help of these sets of static potential, lattice and
analytic, $V_1^{(\rm lat)}$ and $V_1^{(\rm anal)}$, we calculate
the bound states of charmonia, bottomonia, strange quarkonia and
glueballs and compare it with existing lattice data.

Having calculated possible bound states, we discuss their possible
role in the thermodynamics, and in particular, the effective
masses of quarks and gluons and the contribution to the free and
internal energy with the aim to explain the difference
$(\varepsilon - 3p)$ above $T_c$ and ratio
$\varepsilon/\varepsilon_{SB}$.

The paper is organized as follows. In the next section we
introduce the analytic gluelump expression for $D_1^{(E)}$ and the
static potential derived from it.  In section 3 the analytic form
of $D_1^{(E)}$ is tested with existing lattice data in the
interval $1.007 T_c \leq T \leq 1.26 T_c$ and  in section 4 the
static potential is derived from the lattice data. In section 5 we
write the Hamiltonian for heavy and light  quarkonia, glueballs
and  quark-gluon states. In section 6 numerical results for  bound
states in concrete systems are considered  and  discussed in
comparison to lattice data. The possible role of calculated
objects in the thermodynamics of quark-gluon plasma is envisaged,
and the suppression factor for the colored states is derived.

\section{Analytic form of field correlators and the $Q\bar Q$ static potential
at $T=0$}

We start with the standard definitions of the field correlator
functions $D(x)$ and $D_1(x)$, defined as  in \cite{4},  which are
measured in \cite{6,7} $$ \frac{g^2}{N_c}\lan tr_f (F_{\mu\nu} (x)
\Phi (x,y) F_{\lambda\sigma} (y) \Phi (y,x))\ran \equiv
D_{\mu\nu,\lambda\sigma} (x,y)= $$ \be=
(\delta_{\mu\lambda}\delta_{\nu\sigma}-\delta_{\mu\sigma}\delta_{\nu\lambda})
D(x-y) +\frac12\left( \frac{\partial}{\partial x_\mu}
h_\lambda\delta_{\nu\sigma} +perm.\right) D_1(x-y)\label{1}\ee
where $h_\mu=x_\mu-y_\mu,~~\Phi(x,y)=P\exp ig \int^x_y A_\mu d
z_\mu$ and $tr_f$ is the trace in the fundamental representation.
Our final aim in this section will be to connect $D(x), D_1(x) $
to the gluelump Green's functions.  These Green's functions are
however not accessible for direct  analytic calculation and to
proceed one needs to use Background Field Formalism (BFF)
\cite{20}, where the notions of valence gluon field $a_\mu$ and
background field $B_\mu$ are introduced, so that total gluonic
field $A_\mu$
 is written as
\be
A_\mu=B_\mu+a_\mu\label{3} \ee

Using (\ref{3}) one can write the total field operator
$F_{\mu\nu}(x)$ as follows $$ F_{\mu\nu} (x) =\partial_\mu
A_\nu-\partial_\nu A_\mu-ig [A_\mu,A_\nu]=$$
\be
=\partial_\mu(a_\nu+B_\nu)-\partial_\nu(a_\mu+B_\mu)
-ig[a_\mu+B_\mu,a_\nu+B_\nu]=
\label{4}
\ee
$$=\hat D_\mu a_\nu-\hat D_\nu a_\mu-ig[a_\mu,a_\nu]+ F^{(B)}_{\mu\nu}.$$
Here the term, $F_{\mu\nu}^{(B)}$ contains
only the field $B_\mu^b$.  To proceed one can use the idea
\cite{21} of the background field $B_\mu^{(b)}$ as the collective
field of all gluons with color indices $b$, with $b$ occupying
most of indices from 1 to $N_c^2-1$, while the color index $"a"$
of $a_\mu^a$ having only few values. The physical idea, realized
in \cite{19,21} is that after averaging over all fields
$B_\mu^{(b)}$ in the vacuum, the resulting interaction for the
valence gluon is diagonal in index $"a"$ and has the form of the
white adjoint string. It is clear that when one averages over
field $a_\mu^a $ and sums finally over all color indices $a$, one
actually exploits all the fields with color indices from
$F^{(B)}_{\mu\nu}$, so that the term $F_{\mu\nu}^{(B)}$ can be
omitted, if summing over all $a$ is presumed to be done at the end
of calculation. In this section we shall concentrate on the first
two terms on the r.h.s. of (\ref{4}), which yield $D_1(x)$.

Assuming the background Feynman gauge, $ D_\mu a_\mu= 0,$ we shall
define now the gluelump Green's function as
\be
G_{\mu\nu} (x,y) =\lan tr_a a_\mu(x) \hat \Phi(x,y)
a_\nu(y)\ran.\label{5}\ee

We ~~ can~~ also ~~ write~~  for~~ the~~ gluelump ~~Green's~~~
function ~~~ (\ref{5})) \\ $G_{\mu\nu} (x,y) = \delta_{\mu\nu} f
((x-y)^2)$ in the limit of vanishing gluon spin-dependent
interaction, see discussion below.

As a result one obtains from (\ref{4},\ref{5}) the following
connection of $D^{(0)}$ and $f((x-y)^2)$ (see \cite{19} for
details of derivation) \be
 D^{(0)}_{4i,4k} (x,y) = \frac{g^2}{2N_c^2} \left \{\frac{\partial}{\partial
 x_4} \frac{\partial}{\partial y_4} \delta_{ik} f((x-y)^2)
  + \frac{\partial}{\partial x_i} \frac{\partial}{\partial y_k}
  f((x-y)^2)\right\}  ,
\label{11}\ee
 on the other hand using (1) with  $h_\mu\equiv x_\mu-y_\mu$ one
 can express $D^{(0)}_{4i,4k}$ through $D_1$ as
 \be
 D_{4i,4k}^{(0)} (h) = \delta_{ik} D(h) +\frac12 \left(
 \frac{\partial}{\partial x_4} h_4 D_1 \delta_{ik} +
 \frac{\partial}{\partial x_i} h_k D_1\right) \label{12}\ee
 and for $h_i=0, i=1,2,3, h_4\neq 0$ one obtains
 \be
 D_1 (x) =- \frac{2g^2}{N^2_c}\frac{df(x^2)}{dx^2},~~G_{\mu\nu} (x,y)
 = \delta_{\mu\nu} f((x-y)^2) \label{13}\ee

To obtain information about the gluelump Green's function
$G_{\mu\nu}$ one can use the   path-integral representation of
$G_{\mu\nu}(x,y)$ in the Fock-Feynman-Schwinger (FFS) formalism
(see \cite{22} for reviews and original references), which was
exploited for gluelump Green's function in \cite{23} \be
G_{\mu\nu} (x,y) = tr_a \int^\infty_0 ds  (Dz)_{xy} e^{-K}\lan
W^{(F)}_{\mu\nu} (C_{xy})\ran\label{16}\ee where $K= \frac14
\int^s_0 \left( \frac{dz_\mu}{d\tau}\right)^2 d\tau$ and \be
W^F_{\mu\nu}(C_{xy}) =PP_F\left\{\exp (ig \int A_\lambda
dz_\lambda ) \exp (2 ig \int^s_0 d\tau \hat F_{\sigma \rho}
(z(\tau)))\right\}_{\mu\nu}\label{17}\ee
 and the closed contour $C_{xy}$ is formed by the straight line
 from $y$ to $x$ due to the heavy adjoint source Green's function
 and the path of the valence gluon $a_\mu$ from $x$ to $y$.
 Note that the nontrivial $\{\mu\nu\}$ dependence of the r.h.s. of
 (\ref{17}) occurs only due to the $\hat F_{\nu\rho}$; expanding in
 powers of this term, one has $W_{\mu\nu}^F= W^{(0)}
 \delta_{\mu\nu} +  W^{(1)}\hat F_{\mu\nu}
 +...$  In what follows we shall neglect $W^{(n)} $ with $n\geq1$
 since these terms correspond to the effect of spin-dependent
 forces in gluelump, which are relatively small and were accounted
 for in \cite{23}.

 Neglecting in $G_{\mu\nu}$ gluon fields altogether we obtain the
 perturbative result, $G_{\mu\nu}  \to G^{(0)}_{\mu\nu}$
 \be
 G_{\mu\nu}^{(0)} (x,y) =\frac{N_c(N^2_c-1)
 \delta_{\mu\nu}}{4\pi^2(x-y)^2}.\label{18}\ee
This is the leading term in the expansion of $G_{\mu\nu}$ at small
$|x-y|$, while the next  order  term is found in  \cite{19} to be
\be D_1(x) = \frac{4 C_2\alpha_s}{\pi} \left\{ \frac{1}{x^4} +
\frac{\pi^2G_2}{24 N_c} +...\right\}, \label{22b}\ee where $G_2$
is the standard gluonic condensate \cite{24} \be D(0)+D_1(0)
=\frac{g^2}{12 N_c} tr F^2(0) =\frac{\pi^2}{18} G_2.\label{23b}\ee
One can prove  consistency of the  resulting $D_1(x)$ as follows.

 Checking  constant term in (\ref{22b}), one can compare
$D_1(0)$ on the l.h.s. of (\ref{22b}) with the r.h.s., $ D_1(0) =
\frac{\alpha_s C_2}{\pi}\cdot \frac{\pi^2}{18} G_2 =
\frac{\alpha_s C_2}{\pi} (D(0)+D_1(0))$, where $D(0)+D_1(0)$ on
the r.h.s. of (\ref{22b}) are defined by the gluon condensate, Eq.
(\ref{23b}). Since $\frac{ \alpha_s C_2}{\pi} \la 1$, this
estimate of $D_1(0)$ is reasonable and suggests that for $\alpha_s
=0.4$ the magnitude of $D_1(0)$ is 0.2 $D(0)$.

This ratio is in agreement with the lattice calculations in
\cite{6}.

 In another check  one considers the singular term, $D_1^{sing} (x)
=\frac{4C_2\alpha_s}{\pi x^4}$ and inserts it in the static $Q\bar
Q$ potential. The static $Q\bar Q$ potential can be expressed
through $D$ and $D_1$, as was done in \cite{25}:
\beq
V(r) &=& 2r \int^r_0 d\lambda \int^\infty_0 d\nu D(\lambda, \nu) + \int^r_0
\lambda d \lambda \int^\infty_0 d\nu [-2 D(\lambda, \nu) + D_1
(\lambda, \nu) ]
\nonumber \\
&\equiv& V_D(r) +V_1(r) .
\label{24b}
\eeq
Inserting in (\ref{24b}) the perturbative part of $D_1$ from
(\ref{22b}) one obtains the standard color Coulomb potential
$V_C(r)=- \frac{4\alpha_s}{3 r}$, thus checking the correct
normalization of $D_1(x)$.

Another form of $G_{ik} (x,y)$ is available at all distances and
practically important at large $|x-y|$, namely \be G_{ik} (x,y) =
N_c (N^2_c-1)\sum^\infty_{n=0} \Psi^{(i)}_n(0) \Psi^{(k)^+}_n (0)
e^{-M_n|x-y|}\label{20}\ee where $\Psi_n^{(i)}(x), M_n$ are
eigenfunction and eigenvalue of the gluelump Hamiltonian,
\cite{23}, details are given in Appendices 1,2,3,4 of ref.
\cite{19}.

For $\Psi_n^{(\mu)}(0)$ one can use the known  equation, which is
obtained from the eigenfunctions $\Psi_n$ of $H_0$ through the
connection  \be
 \Psi^{(\mu)}_n=
\frac{e_\mu}{\sqrt{2\mu}}\psi_n,~~ \left
(\Psi_n^{(\mu)}(0)\right)^2=\frac{\sigma_{adj}}{4\pi}.\label{24}
\ee

Inserting (\ref{24}) into (\ref{20}) one obtains
\be
G_{\mu\nu} (x,y)\approx  N_c(N^2_c-1) \sum^\infty_{n=0}
\delta_{\mu\nu}\frac{(\sigma_{adj})}{4\pi}e^{-M_n|x-y|}.\label{25}\ee
It is clear that for $x\to y$ the sum in (\ref{25}) diverges and
one should use instead of (\ref{25}) the perturbative answer
(\ref{18}). For large $|x-y|$ one can keep in (\ref{25}) only the
terms with the lowest mass, i.e. for the color electric gluelump
state $1^{--}$, which obtains for spacial $\mu,\nu=i,k$

Thus one gets \be G_{ik}\left|_{|x-y|\to \infty} \right. \approx
(N^2_c-1)\frac{N_c\sigma_{adj}}{4\pi} \delta_{ik}
e^{-M_0|x-y|}.\label{26}\ee The eigenvalue $M_0$ was found in
\cite{23} to be $M_0\cong(1.49\div 0.98)$ GeV for $\sigma_f=0.18 $
GeV$^2$ and $\bar\alpha_s=(0\div 0.195)$. This should be compared
with the value $M_0=1.5 \pm 0.4$ GeV, obtained from the QCD sum
rules in \cite{26} and with the lattice value $M_0\approx 1$ GeV
obtained in \cite{6,7} from asymptotic of $D_1$.

Using (\ref{13}) one can define from (\ref{26}) the
nonperturbative part  of $D_1$, which is valid at large $|x|$, \be
D_1^{(nonp)} (x) = \frac{C_2(f)\alpha_s 2M_0
\sigma_{adj}}{\sqrt{x^2}} e^{-M_0|x|},~~ C_2(f)
=\frac{N^2_c-1}{2N_c}\label{32}\ee and the total $D_1$ due to
(\ref{13}) and (\ref{18}) can be represented as \be D_1(x)
=\frac{4C_2(f)\alpha_s}{\pi x^4}e^{-\gamma|x|} + D_1^{(nonp)}
(x)\label{33}\ee and $\gamma$ plays the role of screening length
of gluon.

 One can easily see that insertion of (\ref{13}) into $V_1(R)$ allows to
 express $V_1(R)$ in terms of $f(x)$,
 \be
 V_1(R) =-\frac{g^2}{N_c^2}\int^\infty_0 d\nu (f(R^2+\nu^2) -f
 (\nu^2)).\label{g29}\ee

 For the purely perturbative $f(x^2)$,
 \be f=f^{(0)} (x^2) =\frac{N_c(N_c^2-1)}{4\pi^2 x^2}\label{g30}\ee
 one has from (\ref{g29}) the standard lowest order Coulomb interaction
 \be
 V_1(R) =-\frac{C_2(f)\alpha_s}{R}.\label{g31}\ee

 Note that the last term in this case, $V_1(\infty) \equiv
 \frac{g^2}{N^2_c}\int^\infty_0 f(\nu^2) d\nu$ is diverging at
 small $\nu$, which is connected to the well-known perimeter
 divergence of fixed-contour Wilson or Polyakov lines, which is
 renormalized on the lattice \cite{12,16},
 subtracting short-distance part of potential at $T=0$.
In what follows we shall concentrate on the nonperturbative part
of $V_1(R)$ representing it  as $V_1(R)=V_1^{(pert)} (R) +
V_1^{(np)} (R)$, where $V_1^{pert}$ can be treated as the screened
Coulomb potential, so that renormalization of the Wilson
(Polyakov) lines would affect only $V_1^{pert}(R)$.

 One can easily see in (\ref{g29}) that $V_1(\infty)$ is actually the
 sum of  self-energy parts of quark and antiquark, occurring due to
 the gluelump exchange with time interval $\nu$.

\section{Analytic form $vs$ lattice data at $T>T_c$}

\noindent A detailed study by numerical simulations on a lattice
of the behaviour of the gauge--invariant two--point correlation
functions of the gauge--field strengths across the deconfinement
phase transition ($T \sim T_c$), both for the pure--gauge $SU(3)$
theory and for full QCD with two flavours, has been performed in
Ref. \cite{DDM03}. {\it Quenched} data published in \cite{DDM03}
agree within errors with previous determinations \cite{npb97}
(obtained on a $16^3 \times 4$ lattice, in a range of distances
from $0.4$ to 1 fm approximately) but have been obtained on a
larger lattice ($32^3 \times 6$) and with much higher statistics.

For the benefit of the reader (and essentially for defining the
notations used in the rest of the paper) we report here some
technical details about the lattice determination of the
correlators in \cite{DDM03}. To simulate the system at finite
temperature, a lattice is used of spatial extent $N_\sigma \gg
N_\tau$, $N_\tau$ being the temporal extent, with periodic
boundary conditions for gluons and antiperiodic boundary
conditions for fermions in the temporal direction. The temperature
$T$ corresponding to a given value of $\beta = 2N_c/g^2$ is given
by \be N_\tau \cdot a = {1 \over T} ~, \label{ntau} \ee where $a$
is the lattice spacing. In the {\it quenched} case $a$ only
depends on the coupling $\beta$ and, from renormalization group
arguments, \be a(\beta) = {1\over\Lambda_L} f(\beta) ~,
\label{abeta} \ee where $f(\beta)$ is the so--called {\it scaling
function} and $\Lambda_L$ is the scale parameter of QCD in the
lattice regularization scheme. At large enough $\beta$, $f(\beta)$
is given by the usual two--loop expression: \be f(\beta) =
\left({8\over33}\,\pi^2\beta\right) ^{ 51/121 }
\exp\left(-{4\over33}\pi^2\beta\right) \left[1+{\cal
O}(1/\beta)\right] ~, \label{2-loop} \ee for gauge group $SU(3)$
and in the absence of quarks. The expression (\ref{2-loop}) can
also be used in a small enough interval of $\beta$'s lower than
the asymptotic scaling region, and then $\Lambda_L$ is an
effective scale depending on the position of the interval
considered. The lattice used in Ref. \cite{DDM03} for the {\it
quenched} case was a $32^3 \times 6$ (in our notation, $N_\sigma =
32$ and $N_\tau = 6$) and the critical temperature $T_c$ for such
a lattice corresponds to $\beta_c \simeq 5.8938$ \cite{Boyd96}.
The range of values of $\beta$'s considered in Ref. \cite{DDM03}
goes from $\beta = 5.85$ to $\beta = 6.10$ and in this interval
the effective scale $\Lambda_L$ is about 4.9 MeV.

At finite temperature ($N_\sigma \gg N_\tau$) the $O(4)$
space--time symmetry is broken down to the spatial $O(3)$ symmetry
and the bilocal correlators are expressed in terms of five
independent functions \cite{1}-\cite{3}, \cite{6,7}. Two of them
are needed to describe the electric--electric correlations: \beq
\lefteqn{ {g^2 \over N_c} \langle tr_f [ E_i (x) \Phi(x,y) E_k (y)
\Phi(y,x) ]
\rangle } \nonumber \\
& & = \delta_{ik} \left[ {D}^E + {D}_1^E + u_4^2 {\partial {D}_1^E
\over \partial u_4^2} \right] + u_i u_k {\partial {D}_1^E \over
\partial \vec{u}^2} ~, \label{EE} \eeq where $E_i = F_{i4}$ is the
electric field operator and $u_\mu = x_\mu - y_\mu$, [$\vec{u}^2 =
(\vec{x} - \vec{y})^2$].

Two further functions are needed for the magnetic--magnetic
correlations: \beq \lefteqn{ {g^2 \over N_c} \langle tr_f [ H_i
(x) \Phi(x,y) H_k (y) \Phi(y,x) ]
\rangle } \nonumber \\
& & = \delta_{ik} \left[ {D}^H + {D}_1^H + \vec{u}^2 {\partial
{D}_1^H \over \partial \vec{u}^2} \right] - u_i u_k {\partial
{D}_1^H \over \partial \vec{u}^2} ~, \label{HH} \eeq where $H_k =
{1 \over 2} \varepsilon_{ijk} F_{ij}$ is the magnetic field
operator.

Finally, one more function is necessary to describe the mixed
electric--magnetic correlations: \be {g^2 \over N_c} \langle tr_f
[ E_i (x) \Phi(x,y) H_k (y) \Phi(y,x) ] \rangle = -{1 \over 2}
\varepsilon_{ikn} u_n {\partial {D}_1^{HE} \over \partial u_4} ~.
\label{EH} \ee In Eqs. (\ref{EE}), (\ref{HH}) and (\ref{EH}), the
five quantities ${D}^E$, ${D}_1^E$, ${D}^H$, ${D}_1^H$ and
${D}_1^{HE}$ are all functions of $\vec{u}^2$, due to rotational
invariance, and of $u_4^2$, due to time--reversal invariance.

The following four quantities have been determined in
\cite{DDM03}\footnote{The definition of the gauge--invariant
field--strength correlation function $D_{\mu\nu,\lambda\sigma}$
adopted in \cite{DDM03} differs from the one given in this paper
by the absence of the multiplicative factor $1/N_c$ on the
left-hand side of Eq. (\ref{1}). Therefore all functions $D^E$,
$D_1^E, \ldots$ used in this paper are {\it smaller} then the
corresponding functions used in \cite{DDM03} by a factor $N_c =
3$.} \beq {D}_\parallel^E (\vec{u}^2,0) &\equiv&
 {D}^E (\vec{u}^2,0) + {D}_1^E (\vec{u}^2,0) +
\vec{u}^2 {\partial{D}_1^E \over \partial \vec{u}^2} (\vec{u}^2,0)
~;
\nonumber \\
{D}_\perp^E (\vec{u}^2,0) &\equiv& {D}^E (\vec{u}^2,0) + {D}_1^E
(\vec{u}^2,0) ~; \label{corr-E}
\\
{D}_\parallel^H (\vec{u}^2,0) &\equiv&
 {D}^H (\vec{u}^2,0) + {D}_1^H (\vec{u}^2,0)
+ \vec{u}^2 {\partial{D}_1^H \over \partial \vec{u}^2}
(\vec{u}^2,0) ~;
\nonumber \\
{D}_\perp^H (\vec{u}^2,0) &\equiv& {D}^H (\vec{u}^2,0) + {D}_1^H
(\vec{u}^2,0) ~, \label{corr-H} \eeq by measuring appropriate
linear superpositions of the correlators (\ref{EE}) and (\ref{HH})
at equal times ($u_4 = 0$). Concerning the mixed
electric--magnetic correlator of Eq. (\ref{EH}), it vanishes both
at zero temperature and at finite temperature, when computed at
equal times ($u_4 = 0$), as a consequence of the invariance of the
theory under time reversal.

The results found in Ref. \cite{DDM03}, both for the {\it
quenched} and the full--QCD case, are in agreement with those
already found in Ref. \cite{npb97} and can be summarized as
follows:
\begin{itemize}
\item[(1)] In the confined phase ($T < T_c$), up to temperatures very
near to $T_c$, the correlators, both the electric--electric type
(\ref{EE}) and the magnetic--magnetic type (\ref{HH}), are nearly
equal to the correlators at zero temperature: in other words,
${D}^E \simeq {D}^H \simeq {D}$ and ${D}_1^E \simeq {D}_1^H \simeq
{D}_1$ for $T < T_c$.
\item[(2)] Immediately above $T_c$, the electric--electric correlators
(\ref{corr-E}) have a clear drop, while the magnetic--magnetic
correlators (\ref{corr-H}) stay almost unchanged, or show a slight
increase.
\end{itemize}
(For the {\it quenched} theory the behaviour of ${D}_\parallel^E$
and ${D}_\perp^E$ is shown in Figs. 1 and 2 of Ref. \cite{DDM03}
respectively, at different values of $T/T_c$ with the physical
distance in the range from $\sim 0.25$ fm up to $\sim 1.25$ fm.
The analogous behaviour for ${D}_\parallel^H$ and ${D}_\perp^H$ is
shown in Figs. 3 and 4 of Ref. \cite{DDM03}.)

Moreover, a best--fit analysis of the data has been performed in
\cite{DDM03}, both for the {\it quenched} and the full--QCD case,
with functions for $D$ and $D_1$ having a perturbative term
$a/x^4$ (here $x = |\vec{u}|$) plus an exponential
non--perturbative term $A \exp(-\mu x)$: this analysis has shown
explicitly that the electric gluon condensate drops to zero at the
deconfining phase transition.

In this section, inspired by the results found in \cite{19} and
partially reported in the previous sections, see Eqs. (\ref{32})
and (\ref{33}), we shall present the results obtained by
performing alternative best fits to the {\it quenched} lattice
data for the electric correlators (\ref{corr-E}) at temperatures
$T$ above the critical temperature $T_c$ (and at equal times,
i.e., $u_4 = 0$), with the functions: \be {D}^E(x) = {a \over x^4}
~~,~~ {D}^E_1(x) = {B \over x} {\rm e}^{-M x} + {b \over x^4}  ~,
\label{fit-E} \ee where, of course, all the coefficients must be
considered as functions of the physical temperature $T$. According
to the results found in Refs. \cite{DDM03,npb97}, the function
$D^E$ is taken to be purely perturbative (so behaving as $1/x^4$)
in the deconfined phase ($T>T_c$). Indeed, from the conclusions of
Refs. \cite{1,2,3}, one expects that the non--perturbative part of
${D}^E$ is related to the (temporal) string tension and should
have a drop just above the deconfinement critical temperature
$T_c$. In other words, the non--perturbative part of ${D}^E$ is
expected to be a kind of order parameter for confinement and this
is fully confirmed by the results found in Refs.
\cite{DDM03,npb97}. On the contrary, ${D}^E_1$ does not contribute
to the area law of the temporal Wilson loop and we use for this
function a parametrization derived from Eqs. (\ref{32}) and
(\ref{33}), consisting in a sum of a non--perturbative term $B/x
~\exp(-Mx)$ plus a perturbative term $b/x^4$. (As already noticed
in \cite{DDM03}, the perturbative coefficients $a$ and $b$ are
regularization--scheme dependent. In Eq. (\ref{fit-E}) we refer to
the lattice regularization scheme; other schemes could give
different values \cite{Jamin98}. We will comment again on this
question in the next section.)

The best fit has been performed to the data for both
(perpendicular and parallel) electric correlators (\ref{corr-E})
for distances from 3 up to 6--7 lattice spacings (corresponding
approximately to the range of physical distances $0.3 \div 0.7$
fm), i.e., for those distances where we have data for {\it both}
parallel and perpendicular electric correlators at all
temperatures (see Figs. 1 and 2 in Ref. \cite{DDM03}).

After having tried many fits with all the parameters free for each
given temperature $T$, which were rather unstable and not strictly
conclusive, we have taken the mass $M$ of the non--perturbative
term of $D_1^E$ in (\ref{fit-E}) and the perturbative coefficients
$a$, $b$ to be temperature independent (at least in the range of
temperatures that we are considering): indeed, this fact, i.e.,
the temperature independence (in the short range of $T$
considered) of the perturbative coefficients and of the
correlation length of the non--perturbative terms, was first
suggested and confirmed in Ref. \cite{DDM03}.\footnote{We remind
again, however, that the non--perturbative terms for the functions
$D$ and $D_1$ in Ref. \cite{DDM03} were taken to be exponentials,
$A \exp(-\mu x)$, with the same correlation length $\lambda =
1/\mu$ for all functions.}

In Table I we report the results obtained by fitting
simultaneously all the data for the perpendicular electric
correlator ${D}_\perp^E$ [see Eq. (\ref{corr-E})] at temperatures
$T>T_c$ with the functions (\ref{fit-E}), where only the
non--perturbative coefficient $B$ is considered to be temperature
dependent. The value of the mass $M$ comes out to be about 1 GeV
and the non--perturbative coefficient $B$ drops rapidly to zero
(within the errors) going from $T/T_c=1.007$ to $T/T_c=1.261$. The
value of the perturbative coefficient $a+b$ is perfectly
consistent (within the errors) with the value $0.90(3)/N_c$ found
in \cite{DDM03}. In Table II we report the results obtained by a
best fit similar to the previous one, but fixing the value of the
perturbative coefficient $a+b$ to the value $0.90/N_c = 0.30$
found in \cite{DDM03}.

Finally, we have performed a best fit to all the values for the
difference \be {D}_\perp^E (x) - {D}_\parallel^E (x) = -{x \over
2} {\partial{D}_1^E \over \partial x} (x) \label{diff-ele} \ee
between the two electric correlators (\ref{corr-E}) at $T>T_c$.
The results are reported in Table III and, within the very large
errors, they roughly agree with those obtained in the two previous
best fits. Let us observe, in particular, that the value of the
perturbative coefficient $b$ agrees, within the errors, with the
value $0.35(1)/N_c$ found in \cite{DDM03}. If we repeat the best
fit by fixing the value of the perturbative coefficient $b$ to the
value $0.35/N_c \simeq 0.12$ found in \cite{DDM03}, we obtained
the results reported in Table IV.

The values of $\chi^2/N$ for the best fits considered are
satisfactory and we can conclude that the functions (\ref{fit-E}),
inspired by Eqs. (\ref{32}) and (\ref{33}), represent a reasonable
parametrization of the correlators in the deconfined phase.

\section{Static potential from lattice correlator\\ measurements}

\noindent In this section we use the results from the best fits to
the lattice data for the electric correlators (\ref{corr-E}) above
$T_c$ with the functions (\ref{fit-E}), that we have obtained in
the previous section, for deriving the $Q\bar{Q}$ static potential
$V_1(R,T)$ in the deconfined phase.

According to the results found in \cite{17}, and partially
reported in the previous sections, the expressions (\ref{fit-E})
for the electric correlation functions $D^E$ and $D_1^E$ imply for
the perturbative and the non--perturbative part of the static
potential at $T>T_c$, \be V_1(R,T) = V_1^{(pert)}(R,T) +
V_1^{(np)}(R,T) ~, \label{V1tot} \ee the following approximate
expressions: \be V_1^{(pert)}(R,T) = -{\pi b \over 4 R} \left[ 1 -
{2 \over \pi} {\rm arctan}(RT) - {RT \over \pi} \ln \left( 1 + {1
\over (RT)^2} \right) \right] ~, \label{V1pert} \ee (satisfying
the condition $V_1^{(pert)}(\infty) = 0$) and: \beq
V_1^{(np)}(R,T) & = & V_1^{(np)}(\infty) - {B \over M^2} \left[
K_1(MR)~MR - {T \over M} e^{-MR} (1 + MR) \right] ~,
\nonumber \\
V_1^{(np)}(\infty) & = & {B \over M^2} \left[ 1 - {T \over M}
\left( 1 - e^{-M/T} \right) \right] ~, \label{V1np} \eeq where
$K_1(x)$ is the modified Bessel function.

As already observed in the previous section, standard perturbation
theory provides the following estimate for the coefficient $b$
[compare the parametrization (\ref{fit-E}) for $D_1^E$ with the
expressions (\ref{32}) and (\ref{33})]: \be b = {4 \alpha_s C_2(f)
\over \pi} + {\cal O}(\alpha_s^2) ~, ~~~~ C_2(f) = {N_c^2 - 1
\over 2N_c} ~, \label{b-pert} \ee and the first
(zero--temperature) term in the right--hand side of
(\ref{V1pert}), i.e., $V_1^{(pert)}(R,T=0) = -\pi b/(4R)$, is
nothing but the standard Coulomb potential written in Eq.
(\ref{g31}).

When extracting the parameter $b$ from lattice measurements of the
correlators, as we have done in the previous section, the coupling
constant $\alpha_s$ in (\ref{b-pert}) must be identified with the
{\it bare} lattice coupling constant, which, expressed in terms of
$\beta = 2N_c/g^2$, is given by $N_c/(2\pi\beta)$. In our case
($N_c=3$) the range of $\beta$ values goes from $5.9$ to $6.1$,
corresponding to an $\alpha_s \simeq 0.08$. Using this value of
$\alpha_s$, one immediately verifies that the values of $b$
reported in Tables III and IV agree almost perfectly with the
perturbative estimate (\ref{b-pert}).

In Fig. 1 we show the behaviour obtained for the static potential
$V_1(R,T)$, given by Eqs. (\ref{V1tot})--(\ref{V1np}), as a
function of the distance $R$, for different values of the
temperature $T>T_c$, using for $B(T)$ and $M$ the central values
reported in Table I and for the parameter $b$ the perturbative
estimate (\ref{b-pert}), with the {\it bare} lattice coupling
constant $\alpha_s = 3/(2\pi\beta)$, as previously discussed.

The error associated with each curve in Fig. 1 can be estimated by
observing that the large--distance behaviour of the potential is
dominated by the non--perturbative part $V_1^{(np)}(\infty)$ in
Eq. (\ref{V1np}), given approximately by $B/M^2$, where, on the
basis of the results reported in Table I, the value of $M$ is
known with a relative error of $10\%$ and the value of $B$ (for
each $T$) is known with a relative error of at least $30\%$. As a
consequence, the non--perturbative part of the potential can be
derived from the results of Table I with a relative error of at
least: \be {\delta V_1^{(np)} \over V_1^{(np)}} \simeq 50 \% ~.
\ee

\section{ Equations for bound states}

In this section we derive Hamiltonian for different quark-gluon
systems, using static potentials (\ref{33}), derived for the color
singlet $Q\bar Q$ system,  which for finite temperature $T$ has
the form \cite{17},\be V_1^{(Q\bar Q)} r, T) = \int^{1/T}_0 d\nu
(1-\nu T) \int^r_0 \xi d \xi D_1(\sqrt{\xi^2+\nu^2})\label{38}\ee
 We can now generalize the static interaction  to the case, when two
 color object $A$ and $B$ with the Casimir coefficients $C(A),
 C(B)$ combine into common color state $D$, with Casimir $C(D)$.
 The answer is\footnote{The construction is similar to that found
 in \cite{36} for the coefficient $\bar b$ of $V_1^{(Q\bar Q)}(r,T)$ but
 differs in   the presence of the first term proportional to $\bar a$,
 since in \cite{35} constant term was not taken into account.}
 $$~^DV^{(AB)} (r,T) =\frac{1}{2C(f)} \{ C(D) V_1^{(Q\bar Q)}
 (\infty, T) + (C(A) +$$\be+C(B) -C(D))V_1^{(Q\bar Q)}(r,T)\}\equiv \bar a
 V_1^{(Q\bar Q)} \infty, T) + \bar bV_1^{(Q\bar Q)} (r,T).\label{39}\ee
 Here $C(f)=\frac43$ is the Casimir number for the fundamental
 charges. To illustrate (\ref{39})  several examples are considered
 in  Table 5.

 For multicomponent systems one can similarly find,
$$ ~^DV^{(ABC)} (\ver_1,\ver_2, \ver_3, T) = \frac{1}{2C(f)} \{C(D)
V_1^{(Q\bar Q)} (\infty, T) +$$\be +(C_A+C_B+C_C -C_D)\frac13
\sum_{i>j} V_1^{(Q\bar Q)}(\ver_i-\ver_j, T)\}\label{40}\ee in
particular \be (QQQ)_1 : V_1^{(QQQ)} =\frac12 \sum_{i>j}
V_1^{(Q\bar Q)} (\ver_i-\ver_j,T)\label{41}\ee
 \be (QQQ)_{10} : V_{10}^{(QQQ)} =\frac94 V_1^{(Q\bar Q)} (\infty,
 T) -\frac14 \sum_{i>j} V_1^{(Q\bar Q)}
 (\ver_i-\ver_j,T)\label{42}\ee

\be (QQQ)_{8} : V_{8}^{(QQQ)} =\frac98 V_1^{(Q\bar Q)} (\infty,
 T) +\frac18 \sum_{i>j} V_1^{(Q\bar Q)}
 (\ver_i-\ver_j,T)\label{43}\ee

 One can check that  at large  distances all systems consisting of
 $n_Q$ quarks or antiquarks and $n_g$ gluons tend to the constant
 limit, independent of $D$
 $$
 ~^DV^{(n_Q Q, n_g g)} (|\ver_i-\ver_j|\to \infty) = E_Q \cdot n_Q
 +E_g n_g,$$\be  E_Q =\frac12 V_1^{(q\bar Q)}(\infty, T),~~ E_g
 =\frac98 V_1^{(Q\bar Q)} (\infty, T).\label{44}\ee

 Since nonperturbative part of $V_1^{(Q\bar Q)}(r,T)\sim O(r^2),
 r\to 0$ one obtains the lower bound on the nonperturbative part
 of $V_D^{(n_Q Q, n_gg)} (r_{ij}, T)$, which  is
 $\frac{C(D)}{2C(f)} V_1^{(Q\bar Q)} (\infty, T)$.

 As a consequence, one can predict the absence of bound states  in
 some channels, e.g. in $(Q\bar Q)_8, (QQ)_6, (QQQ)_{10}$ etc.

As one application of the general relation (\ref{40})
 we show in Fig.2 the  static potentials $V^{(D)}
 (\ver_1,\ver_2,\ver_3, T) = V^{(D)} (R,T)$ of three static
 fundamental quarks in three different representations $D$: singlet
 $(C(D)=0)$, octet$(C(D)=3)$ and decuplet, $(C(D)=6)$ in the
 symmetric configuration with $r_i=R, i=1,2,3,$ as function of
 $R$.One can see in Fig.2 that all three potentials tend to the
 same limit $\frac32 V_1^{(Q\bar Q)} (R,T)$ at large $R$, in
 accordance with Eq. (\ref{44}), while deviations from asymptotic
 at all distances are proportional to $\left(\frac12, -\frac14,
 \frac18\right)$ for singlet, decuplet and octet respectively, as
 prescribed by Eqs. (\ref{41}-\ref{43}). This is in  agreement with
 lattice calculations of free energies $F_{qqq}^{(D)} (R,T)$
 presented in \cite{32}.

Having constructed static potentials for different systems, we can
now exploit the relativistic Hamiltonian technic, developed in
\cite{36} and  successfully used for mesons, baryons, glueballs
and hybrids in the  confinement phase (see \cite{37} for a
review). This technic does not take into account  chiral degrees
of freedom and is applicable when spin-dependent interaction can
be treated as perturbation. Therefore below we stick to the
Hamiltonian technic of \cite{36} and consider heavy quarkonia, and
baryons, leaving light quarkonia with chiral symmetry restoration
to another publication.

Leaving details of derivation to \cite{36,37}, one can write the
bound-state equation as
\be
H\psi_n = \varepsilon_n \psi_n,
\nonumber
\ee
\be
M_n = \min_\mu \left\{ \sum^{n_Q+n_g}_{i=1}
\left(\frac{m^2_i}{2\mu_i} +\frac{\mu_i}{2} \right)+\varepsilon_n
(\mu_1,...\mu_{n_Q+n_g})+ \bar aV_1^{(Q\bar
Q)}(\infty,T)\right\},
\label{45}
\ee
where we have introduced the
einbein variables $\mu_i$ for quarks and gluons with the
stationary values $\mu_i^{(0)}$ playing the role of the
constituent masses. The Hamiltonian has the form \be
H=H_0+H_S+H_{SE}\label{46}\ee where \be H_0=\sum^{n_Q+n_g}_{i=1}
\frac{\vep^2_i}{2\mu_i} + V_D^{(n_Qq, n_gg)}
(\ver_{ij})\label{47}\ee and $H_S, H_{SE}$ are spin-dependent and
self-energy parts of Hamiltonian defined in terms of field
correlators.

The search for bound states in the $Q\bar Q$ and $QQQ$ systems was
done using two types of interaction $V_1^{(Q\bar Q)}(r,T)$.
\begin{description}
  \item[i)]  The first one based on the analytic representation
  (\ref{32}), (\ref{33}) obtained in the confined region, and
  analytically continued into deconfinement as
  \be V_1^{Q\bar Q} (r,T) = V_1^{(np)} (r,T) +
  V_1^C(r,T)\label{48}\ee
  with
  \be V_1^{(np)} (r,T) = a(T) \int^{1/T}_0 d\nu (1-\nu T) [
  e^{\nu M_0}-e^{-\sqrt{\nu^2+r^2}M_0}]\label{49}\ee
  \be V_1^C(r,T) =- \frac{4\alpha_s}{3r} \varphi
  (r,T),~~\alpha_s=0.3\label{50}\ee
  with the  temperature-modified Coulomb term as in (\ref{V1pert})
  \be
  \varphi(r,T) =\left(1-\frac{2}{\pi} \arctan (r T)
  -\frac{rT}{\pi} \ln \left[1 +\frac{1}{(rT)^2}\right]\right)
  e^{-\gamma r}.\label{51}\ee
  Here the coefficient $a(T)$ is
  \be
  a(t) =a_0 - 0.36 \frac{T-T_c}{T_c},\label{52}\ee
  and the value of $a_0$ coincides with that obtained in
  (\ref{32}) for the confinement  region,
  \be
  a_0=2C_2(f) \alpha_s \sigma_{adj}=0.648~{\rm GeV}^2\label{53}\ee
  while $M_0$ is taken in  one case the same as the lowest gluelump
  mass $M_0\cong 1$ GeV \cite{23}, and in another it was assumed
  to be decreased  in the deconfined region to the value
  $M_0=0.69$ GeV. The value of $\gamma$ was taken at $\gamma=0.2$
  GeV and 0.69 GeV.
  \item[ii)]  The second choice is based on the lattice
  determination of the correlator $D_1^E(x)$, done in \cite{DDM03,npb97}
  and described above in sections 3 and 4,  and in Fig. 1. Here
  one must rescale the resulting $D_1^E(x)$, taking into account
  that lattice coupling $\alpha_s \approx 0.08$. From theoretical
  definition of $D_1^{(nonpep)}(x)$ through the gluelump Green's
  function in (\ref{32}) one can  see, that $D_1^{(nonp)}$ is
  proportional to $\alpha_s$ (with higher order terms proportional to
  $\alpha_s^n, n\geq2$ times  more complex gluelump
  Green's functions, as discussed in \cite{19}).
 Since we are interested in large   distance behaviour of
 $V_1^{(np)} (r,T)$, one should use the infrared-saturated value
 of $\alpha_s(\infty)$, which in the confined region (and
 hopefully at $T\geq T_c$ but close to $T_c$) can be taken from
 analysis of meson spectra \cite{b35} and background perturbation
 theory \cite{20}. According to this the acceptable value is $a_s
 (\infty)\approx 0.6 \div 0.45 $ for $n_f=5\div 0$.

In Fig.1 a rescaling parameter of the lattice defined
 $D_1^{(nonp)}(x) $ and $V_1^{(np)}$  is  taken as  $\xi =
 \frac{\alpha_s(\infty)}{\alpha_s(lattice)}= \frac{0.6}{0.08}
 =7.5$.

 In Fig.1 on the r.h.s. of the figure, the rescaled values of
 $V_1^{(np)}$ are given, and one can find the maximum value of
 $V_1^{(np)}(\infty, T)$ at $T=1.007$ is equal to $\sim 0.53$ GeV.
 These values are in farly good agreement (within the large errors)
 with the lattice measurements of $F_1(r,T)$ for $T>T_c$ reported
 in \cite{16}.

 \end{description}

The two interactions, described in  {\bf i)}
 and {\bf ii)} respectively, and denoted $V_1(I)$ and $V_1(II)$,
 have been used to calculate the bound states of different binary
 systems, both white and colored, as shown in Table 6.

 The mass of the binary state can be computed according to
 (\ref{45}-\ref{47}), with inclusion of the total selfenergy for
 colored bound states,
\be M_i=\bar a_i V_1^{(np)} (\infty, T) + min_{\mu_1,\mu_2} \left(
\frac{m_1^2}{2\mu_1}+ \frac{m^2_2}{2\mu_2}+ \frac{\mu_1+\mu_2}{2}
+\varepsilon^{(i)}(\tilde\mu)\right)\label{54}\ee where the
eigenvalue $\varepsilon^{(i)} (\tilde \mu)$ is a solution of
equation \be \left( \frac{\vep^2}{2\tilde\mu} + \bar b_iV_1^{Q\bar
Q} (r,T) \right) \psi(r) = \varepsilon^{(i)} (\tilde \mu)
\psi(r).\label{55}\ee

Here $\tilde \mu=\frac{\mu_1\mu_2}{\mu_1+\mu_2}$ and $m_1, m_2,
\bar a_i, \bar b_i$ are listed in Table 6.

Several words  about the choice of masses $m_1,m_2$ in Table 6. In
the confined phase at $T=0, m_1, m_2$ are current masses and their
values for charm and bottom quark are assumed to be the same for
$T>T_c$ and correspond to the values  listed in PDG \cite{38} and
used for heavy quarkonium spectra in \cite{39}.
 For light quarks and gluons for $T>T_c$ one should take the
 values either   found from the lattice analysis \cite{40}, or
 from the quasiparticle calculations of quark-gluon plasma \cite{41} (see
 also discussion in \cite{35}). As a result we have chosen some
 averaged effective values of $m_q, m_g$ listed in Table 6.

 Eq.(\ref{55}) has the form of nonrelativistic Schroedinger
 equation, however one immediately realizes inserting (\ref{55}) in
 (\ref{54}), that minimization in $\mu_1,\mu_2$ yields $$
 \min_{\mu_1,\mu_2} \left( \frac{m^2_1}{2\mu_1}
 +\frac{m^2_1}{2\mu_2} +\frac{\mu_1+\mu_2}{2}
 +\frac{\vep^2}{2\tilde \mu}\right)=
 \sqrt{\vep^2+m^2_1}+\sqrt{\vep^2+m^2_2},$$ i.e. one obtains the
 Salpeter equation, which accounts for the relativistic
 kinematics. From this simple exercise (valid for zero orbital
 momentum) one can recognize the physical meaning of einbein
 variables $\mu_1,\mu_2$, namely $\mu_i\sim \lan
 \sqrt{\vep^2+m^2_i}\ran$ which play the role of constituent
 masses of the particles (for more discussion see \cite{36,37}).

 In particular Eqs.(\ref{54},\ref{55}) have been used in \cite{42}
 to calculate glueball masses at $T=0$ with $V_1^{Q\bar Q} =\sigma
 r$, (which appeared in perfect agreement with lattice data
 \cite{43}) where $\mu$ for gluons  in the lowest mass glueball is
 calculated to be  $\mu=0.528$ GeV.

Hence one is justified to apply Eqs. (\ref{54},\ref{55}) both to
heavy quark and light quark and gluon systems listed in Table 6.

\section{Results and discussion}

Eqs. (\ref{54}), (\ref{55})  have been solved numerically  for the
parameters shown in Table 6.

First of all   we have concentrated on the charmonium bound states
for the potential $V_1(I)$  and  considered four combinations of
parameters $M_0,\gamma$ for the analytic  potential
(\ref{48}-{\ref{53}):
\begin{eqnarray}
1)~~ M_0&=&1~{\rm GeV}~,~~ \gamma= 0.69 {\rm~ GeV} \nonumber\\
2)~~ M_0&=&1~{\rm GeV}~,~~ \gamma= 0.2 {\rm~ GeV} \nonumber\\ 3)~~
M_0&=&0.69~{\rm GeV}~,~~ \gamma= 0.69 {\rm~ GeV}\nonumber\\ 4)~~
M_0&=& 0.69~{\rm GeV}~,~~ \gamma= 0.2 {\rm~ GeV}. \label{56}
\end{eqnarray}

Results of numerical calculations   of bound states  masses
$M(c\bar c)$ for the cases 1)-4) are presented in Table VII and
Fig. 3. As expected, the cases 3) and 4) give the deepest  bound
states which keep intact to largest temperatures, $T=1.6 T_c$ and
$T=1.8 T_c$ respectively. For $T=T_c$  charmonium $L=0$ masses lie
in the interval $3.3\div 3.4$ GeV.

To follow in more detail the process of bound state dissolution
with increasing temperature, we plot in Fig.4 the values of
binding energies  $\varepsilon \equiv M(c\bar c)-2m_c-V_1(\infty,
T)$. One can see in Fig.4 that all 4 bound state levels are diving
in the barrier $V_1(\infty, T)$ at some critical value of $T/T_c$,
where the radius of bound states is infinitely increasing, as one
can see from the Table VII.

A similar situation occurs for the potential $V_1(II)$, derived
from the lattice measurements of $D_1^E(x)$ \cite{DDM03,npb97}.
The masses $M(c\bar c)$ are presented in Table VIII for two values
of gluon screening mass   $\gamma, \gamma= M_0 = 1.044 $ GeV and
$\gamma=0.2$ GeV. One can see, that the $c\bar c$ bound  states
dissolve at  lower temperatures, 1.007$T_c$ and 1.065$T_c$
respectively which reflects the fact, that the amplitude of
$V_1(\infty, T)$ drops much faster  with temperature  for the
potential $V_1(II)$ than for the analytic potential $V_1(I)$.

Results of numerical calculations for the bottomonium bound state
mass $M(b\bar b)$ are presented in Fig.5 for the parameter
combinations 2 and 4 from (\ref{56}). As expected, bound states
exist in  a wider temperature region up to $T\cong 2.2 T_c$.

At this point it is possible to compare $M(c\bar c)$ as function
of $T$ to experimental masses at $T=0$ and to the MEM lattice
calculations.

 One can see in Tables VII and VIII that immediately above $T_c$ the
  values of $L=0, M(c\bar c)$ are about 0.3 GeV higher than the
  $J/\psi$ mass of 3.1 GeV and at higher  temperatures $M(c\bar
  c)$ drop below this level. One can see the same type of
  qualitative behaviour in MEM calculations, (e.g. in Fig. 3 of
  \cite{42g} and  Fig.15 of  \cite{13}), however quantitative
  accuracy of MEM masses is  difficult to ascertain.

  Of special interest is the $L=1, c\bar c$ bound  state. In our
  calculations for the most favourable case 4 this bound state
  appears at the threshold only for the interaction increased 2.5
  times (at $T=T_c$), with the mass around 4 GeV. This large
  difference is explained by the small radius $1/M_0\sim 1$
  GeV$^{-1}$ of the potential $V_1(r,T)$. Thus it is not possible
  to have $P$ state $c\bar c$ bound states with the interaction
  $V_1(I)$ (and even more so for $V_1(II))$.
  At $T=0$ the $\chi_{c0}$ state is only 350 MeV above  $J/\psi$,
  which corresponds to larger radius of confining interaction (the
  radius of $\chi_{c0} $ is  around 0.6 fm). In \cite{13,42g}
  the mass of  $\chi_{c0}$ for $ T>T_c$ appears below 3 GeV as wide
  bumps in the MEM analysis. If this can be considered as $L=1$,
  $c\bar c$ bound states,  the very fact that they appear below
  $L=0, c\bar c$ states is difficult to explain. The
  mechanism suggested in \cite{1}, which takes into account the
  Thomas spin-orbit interaction,
  $V_{S)}(r) =- \frac{\sigma_H \veL\veS}{m^2_c r},$
  with spacial string tension  $\sigma_H =\frac12 \int  D^H (x)
  d^2x$, $\sigma_H\geq \sigma_E=0.18$ GeV$^2$ for all $T\geq T_c$,
  yields attractive bound states for
  $J=L+1$ with accumulation point just below $2m_c =2.8$ GeV. (Note
  however that this mechanism acts for $\chi_{c2}$ rather than
  $\chi_{c0}$).
Let us now turn to other states from the Table VI.

The interesting light quarkonia $L=0$ state with thermal quark
masses of 0.4 GeV appears to be bound only if one multiplies
potential $V_1(r,T)$ by the factor 1.33 for the most favorable
parameter set 4. The resulting total mass, as seen from Table 9,
is around 1.7 GeV. This is agreement with lattice MEM data
\cite{8,11}, however one should have in mind, that the approximate
degeneration of $S,V,A,P$ quarkonia states found there, requires
the use  of another relativistic formalism accounting for
restoration of chiral invariance, and will be reported elsewhere.
As it is, the existence of bound light quarkonia in \cite{8,11}
implies that our potentials $V_1(I), V_1(II)$ underestimate  the
actual interaction and should be possibly increased by 30-50\%.
This agrees with derivation of Eq. (\ref{18}), where only the
lowest mass $M_0$ term is kept, while the next,  radial excited
gluelump term  increases $V_1(r,T)$ by roughly 50\%.

Another white state is glueball $(gg)_1$ in Fig. 6   and Table 9.
One can see that its mass is around 2.2 GeV for $T=T_c$, which is
close to the spin-averaged $T=0, L=0$ level \cite{42,43}. Lattice
glueball MEM data exist for $T\leq T_c$ \cite{43g} and are very
welcome for $T>T_c$. Of special interest are the colored bound
states shown in Fig. 6 and Table 9. One  can see that the triplet
$(cc)$, $(cg)$ and octet glueball survive above $T_c$. In
particular the binding energy  of $(cg)_3$ state is small, which
means that each $c$ quark is surrounded by a widespread cloud of
gluons, which increases its entropy and suppresses the quark
Polyakov line.

All data in Fig. 6 and Table 9 are obtained assuming the potential
$V_1(I)$. For the potential $V_1(II)$ results are similar,  but
the bound states disappear at smaller $T$, and in some cases they
do not exist for the nominal amplitude of interaction.

Our $(c\bar c)$ and $(b\bar b)$ bound states found from the
theoretical and lattice field correlators shown in Figs. 3-5 and
Tables VII,VIII and IX can be compared to the calculations based
on phenomenological potentials \cite{44,45}. In \cite{44} the
potential was obtained identifying $V_1(r,T)$ with the free energy
$F_1(r, T)$, measured on the lattice and the resulting  binding
energies $\varepsilon$ for $(c\bar c)$, $(b\bar b)$ systems are
close to our results shown in Tables VII, VIII and Figs. 3-5.

Alberico et al. \cite{45} have fitted the $Q\bar Q$ potential to
the lattice free energy data and found $c\bar c$ and $b\bar b$
bound states in the interval $T_c \leq T\leq T_d$, $T_d/T_c\approx
2 $ for $J/\psi$, which roughly agrees with our results, whereas
our interval for $\Upsilon$ is much shorter.

The behaviour of  glueball masses  is important both from the
point of view of thermodynamics of the phase transition \cite{2,
46}, and for the  role of glueballs in the hadronizing
quark-gluon plasma \cite{47}.

The final comment is on the thermodynamics in the quark-gluon
plasma, taking into account strong interaction and possible white
and colored bound states. An extensive discussion of this topic
was done in \cite{48}-\cite{50}. In particular, it was proposed in
\cite{35,48}, that abundance of colored bound states appearing due
to strong Coulomb potential (the plateau's in $V_1(r,T)$ and the
term $a_iV_1(\infty, T)$ for colored
 states were not discussed in \cite{35,48}) can drastically change
 the plasma thermodynamics.

 However, the presence of self-energy parts  for colored states in
 the form of the term $\bar a_i V_1(\infty, T)$ in the total mass
 (\ref{54}), can modify contribution of colored states to the
 plasma partition function. E.g., each state appears with
 additional Boltzmann factor $\exp \left( -\frac{\bar a_i
 V_1(\infty, T)}{T} \right)$ and for $(cg)_6$ state this
 amounts to the coefficient $\sim 0.01$ for $T\simeq T_c=0.17$
 GeV.

 Therefore the effective number of d.o.f. in plasma may be much
 smaller, than one could expect. This conclusion is  seemingly in
 agreement  with recent studies \cite{51}-\cite{53}.

\vfill\eject

\noindent
\begin{center}
{\bf TABLE CAPTIONS}
\end{center}
\vskip 0.5 cm
\begin{itemize}
\item [\bf Tab.~I.] Results obtained by fitting simultaneously all the
data for the perpendicular electric correlator ${D}_\perp^E$ [see
Eq. (\ref{corr-E})] at temperatures $T>T_c$ with the functions
(\ref{fit-E}), where only the non--perturbative coefficient $B$ is
considered to be temperature dependent.
\bigskip
\item [\bf Tab.~II.] Results obtained by a best fit similar to that of
Table I, but fixing the value of the perturbative coefficient
$a+b$ to the value $0.90/N_c = 0.30$ (marked with an asterisk in
the Table) found in \cite{DDM03}.
\bigskip
\item [\bf Tab.~III.] Results obtained by fitting simultaneously all the
data for the difference between the perpendicular electric
correlator ${D}_\perp^E$ and the parallel electric correlator
${D}_\parallel^E$ [see Eqs. (\ref{corr-E}) and (\ref{diff-ele})]
at temperatures $T>T_c$ with the functions (\ref{fit-E}), where
only the non--perturbative coefficient $B$ is considered to be
temperature dependent.
\bigskip
\item [\bf Tab.~IV.] Results obtained by a best fit similar to that of
Table III, but fixing the value of the perturbative coefficient
$b$ to the value $0.35/N_c \simeq 0.12$ (marked with an asterisk
in the Table) found in \cite{DDM03}.
\bigskip
\item [\bf Tab.~V.]
Parameters of the static potential of binary systems $(Q\bar Q)_D$, $(QQ)_D$,
$(Qg)_D$ and $(gg)_D$ in different color representations $D$.\\
$V_D^{AB} (r,T) =\bar a V_1^{Q\bar Q} (\infty,T)+ \bar
bV_1^{(Q\bar Q)} (r,T)$.
\item [\bf Tab.~VI.]
Parameters of binary systems, used to calculate bound state
energies with potentials $V_1(I) $ and $V_1(II)$. Masses are given
in GeV.
\item [\bf Tab.~VII.]
Masses of $c\bar c$ bound states for  the potential $V_1(I)$ (in GeV)
for four different sets of $M_0, \gamma$ as in Eq. (\ref{56}) as function
of $T/T_c$ -- second column. Mean radii of the bound states in GeV$^{-1}$
-- third column. Binding energies of bound states (in GeV) with respect to
the barrier height $V_1(\infty, T),
~~ \varepsilon \equiv M-2m_c -V_1(\infty, T)$ -- fourth column.

\item [\bf Tab.~VIII.]
The same as in Tab. VII, but for the potential $V_1(II).$

\item [\bf Tab.~IX.]
The same as in Tab. VII  for the  systems $(s\bar s)_1, (c\bar
c)_3, (cg)_3, (gg)_1, (gg)_8$   with parameters given in Tab. VI.

\end{itemize}

\vfill\eject

\vskip 1cm

\centerline{\bf Table I}

\vskip 5mm

\moveright .2 in \vbox{\offinterlineskip \halign{\strut \vrule
\hfil\quad $#$ \hfil \quad & \vrule \hfil\quad $#$ \hfil \quad &
\vrule \hfil\quad $#$ \hfil \quad & \vrule \hfil\quad $#$ \hfil
\quad & \vrule \hfil\quad $#$ \hfil \quad & \vrule \hfil\quad $#$
\hfil \quad \vrule \cr \noalign{\hrule} \beta & T/T_c & B
& M        & a + b   & \chi^2/N \cr
      &       & ({\rm MeV}^3)        & ({\rm MeV}) &      &          \cr
\noalign{\hrule} \noalign{\hrule}
      &       &                      & 1042(91) & 0.29(4) & 0.28 \cr
\noalign{\hrule} 5.90  & 1.007 & 1.09(34) \times 10^8 &          &
&      \cr \noalign{\hrule} 5.92  & 1.030 & 0.62(20) \times 10^8 &
&         &      \cr \noalign{\hrule} 5.95  & 1.065 & 0.41(14)
\times 10^8 &          &         &      \cr \noalign{\hrule} 6.00
& 1.127 & 0.18(12) \times 10^8 &          &         &      \cr
\noalign{\hrule} 6.10  & 1.261 & 0.00(44) \times 10^8 &          &
&      \cr \noalign{\hrule} }}

\vskip 4cm

\centerline{\bf Table II}

\vskip 5mm

\moveright .2 in \vbox{\offinterlineskip \halign{\strut \vrule
\hfil\quad $#$ \hfil \quad & \vrule \hfil\quad $#$ \hfil \quad &
\vrule \hfil\quad $#$ \hfil \quad & \vrule \hfil\quad $#$ \hfil
\quad & \vrule \hfil\quad $#$ \hfil \quad & \vrule \hfil\quad $#$
\hfil \quad \vrule \cr \noalign{\hrule} \beta & T/T_c & B
& M        & a + b   & \chi^2/N \cr
      &       & ({\rm MeV}^3)        & ({\rm MeV}) &      &          \cr
\noalign{\hrule} \noalign{\hrule}
      &       &                      & 1043(100) & 0.30(*) & 0.39 \cr
\noalign{\hrule} 5.90  & 1.007 & 0.99(33) \times 10^8 &          &
&      \cr \noalign{\hrule} 5.92  & 1.030 & 0.53(18) \times 10^8 &
&         &      \cr \noalign{\hrule} 5.95  & 1.065 & 0.32(11)
\times 10^8 &          &         &      \cr \noalign{\hrule} 6.00
& 1.127 & 0.09(9) \times 10^8  &          &         &      \cr
\noalign{\hrule} 6.10  & 1.261 & 0.00(3) \times 10^8  &          &
&      \cr \noalign{\hrule} }}

\vfill\eject

\vskip 1cm

\centerline{\bf Table III}

\vskip 5mm

\moveright .2 in \vbox{\offinterlineskip \halign{\strut \vrule
\hfil\quad $#$ \hfil \quad & \vrule \hfil\quad $#$ \hfil \quad &
\vrule \hfil\quad $#$ \hfil \quad & \vrule \hfil\quad $#$ \hfil
\quad & \vrule \hfil\quad $#$ \hfil \quad & \vrule \hfil\quad $#$
\hfil \quad \vrule \cr \noalign{\hrule} \beta & T/T_c & B
& M        & b       & \chi^2/N \cr
      &       & ({\rm MeV}^3)        & ({\rm MeV}) &      &          \cr
\noalign{\hrule} \noalign{\hrule}
      &       &                      & 887(830) & 0.14(4) & 0.33 \cr
\noalign{\hrule} 5.90  & 1.007 & 0.13(49) \times 10^8 &          &
&      \cr \noalign{\hrule} 5.92  & 1.030 & 0.15(51) \times 10^8 &
&         &      \cr \noalign{\hrule} 5.95  & 1.065 & 0.11(44)
\times 10^8 &          &         &      \cr \noalign{\hrule} 6.00
& 1.127 & 0.06(33) \times 10^8 &          &         &      \cr
\noalign{\hrule} 6.10  & 1.261 & 0.00(28) \times 10^8 &          &
&      \cr \noalign{\hrule} }}

\vskip 4cm

\centerline{\bf Table IV}

\vskip 5mm

\moveright .2 in \vbox{\offinterlineskip \halign{\strut \vrule
\hfil\quad $#$ \hfil \quad & \vrule \hfil\quad $#$ \hfil \quad &
\vrule \hfil\quad $#$ \hfil \quad & \vrule \hfil\quad $#$ \hfil
\quad & \vrule \hfil\quad $#$ \hfil \quad & \vrule \hfil\quad $#$
\hfil \quad \vrule \cr \noalign{\hrule} \beta & T/T_c & B
& M        & b       & \chi^2/N \cr
      &       & ({\rm MeV}^3)        & ({\rm MeV}) &      &          \cr
\noalign{\hrule} \noalign{\hrule}
      &       &                      & 1154(186) & 0.12(*) & 0.33 \cr
\noalign{\hrule} 5.90  & 1.007 & 0.52(25) \times 10^8 &
&         &     \cr \noalign{\hrule} 5.92  & 1.030 & 0.54(25)
\times 10^8 &           &         &     \cr \noalign{\hrule} 5.95
& 1.065 & 0.47(21) \times 10^8 &           &         &     \cr
\noalign{\hrule} 6.00  & 1.127 & 0.37(16) \times 10^8 &
&         &     \cr \noalign{\hrule} 6.10  & 1.261 & 0.30(12)
\times 10^8 &           &         &     \cr \noalign{\hrule} }}

\vfill\eject

\vskip 4cm

\centerline{\bf Table V}

\vskip 5mm

\begin{center}

\begin{tabular}{|c|c|c|c|c|c|c|c|c|}\hline &&&&&&&&\\
system& $(Q\bar Q)_1$&$(Q\bar Q)_8$&$(Q Q)_3$&$(Q Q)_6$&$(Q
g)_3$&$(Q g)_6$&$(gg)_1$&$(gg)_8$\\&&&&&&&&\\ \hline
&&&&&&&&\\
$a$& 0 &$\frac98 $&$\frac12 $&$\frac54 $&$\frac12 $&$\frac54 $&0&
$\frac98 $\\
&&&&&&&&\\ \hline
&&&&&&&&\\
$b$&1 &$-\frac18 $&$\frac12 $&$-\frac14 $&$\frac98 $&$\frac38 $&
$\frac94 $&1\\ &&&&&&&&\\ \hline \end{tabular}

\end{center}

\vskip 2cm

\centerline{\bf Table VI}

\vskip 5mm

\begin{center}

\begin{tabular}{|c|c|c|c|c|c|c|c|c|}\hline &&&&&&&&\\
i& 1&2&3&4&5&6&7&8\\  &&&&&&&&\\ \hline  &&&&&&&&\\ contents&
$(b\bar b)_1$& $(c\bar c)_1$&$(s\bar s)_1$&$(cc)_{\bar
3}$&$(cg)_3$& $(cg)_6$&$(gg)_1$& $(gg)_8$\\  &&&&&&&&\\ \hline
&&&&&&&&\\ $m_1^{(i)}$&4.8&1.4&0.4&1.4&1.4&1.4&0.5&0.5\\
&&&&&&&&\\ \hline  &&&&&&&&\\
$m_2^{(i)}$&4.8&1.4&0.4&1.4&0.5&0.5&0.5&0.5\\  &&&&&&&&\\ \hline  &&&&&&&&\\
$\bar a_i$&0&0&0&$\frac12$&$\frac12$&$\frac54$&0&$\frac98$\\  &&&&&&&&\\
\hline  &&&&&&&&\\
$\bar b_i$&1&1&1&$\frac12$&$\frac98$&$\frac38$&$\frac94$&1\\  &&&&&&&&\\
\hline
\end{tabular}

\end{center}

\vskip 2cm

\centerline{\bf Table VII}

\vskip 5mm

\begin{center}

\begin{tabular}{|c|c|c|c|c|}\hline
  &$ T/Tc $&$      M $&$           <r>  $&$    \varepsilon
  $\\  \hline
1)& \multicolumn{4}{|c|}{$  M_0=1.0 ~$ ~~~$  \gamma=0.69$}\\
&  1.0 & 3.33&   6.62& -0.0063\\
&  1.1 & 3.28&   11.16& -0.0018\\ \hline

2)& \multicolumn{4}{|c|}{$  M_0=1.0$ ~~~~$  \gamma=0.2$}\\
&  1.0&  3.30 &  3.78& -0.031\\
&  1.1 & 3.26 &  4.41 &-0.020\\
&  1.2&  3.21 &  5.40 &-0.012 \\
  &1.3 & 3.17 &  7.14 &-0.006   \\
 & 1.4 & 3.12  & 11.11 &-0.002      \\
 & 1.5 & 3.08   &23.07 &-0.0002        \\
 \hline
3)& \multicolumn{4}{|c|}{ $ M_0=0.69$ ~~~~$  \gamma=0.69$}\\
  &1.0 & 3.40  & 2.77& -0.107   \\
&  1.1 & 3.35  & 3.06& -0.077    \\
&  1.2 & 3.30  & 3.45& -0.053     \\
 & 1.3 & 3.25  & 4.02& -0.033     \\
 & 1.4 & 3.20  & 4.91& -0.018       \\
  &1.5 & 3.15  & 6.67& -0.008       \\
  &1.6 & 3.09  & 12.13& -0.002
  \\\hline
4)& \multicolumn{4}{|c|}{ $ M_0=0.69$~~~~ $  \gamma=0.2$}\\
&  1.0 & 3.36 &  2.48&-0.150  \\
&  1.1 & 3.31 &  2.69&-0.115\\
&  1.2 & 3.27 &  2.95&-0.086\\
&  1.3 & 3.22 &  3.29&-0.061 \\
&  1.4 & 3.18 &  3.77&-0.040  \\
&  1.5 & 3.13 &  4.47&-0.024  \\
&  1.6 & 3.08 &  5.67&-0.012   \\
&  1.7 & 3.03 &  8.30&-0.004   \\
&  1.8 & 2.98 &  18.53 &-0.0005    \\ \hline

  \end{tabular}

  \end{center}

\newpage

   \vskip 2cm

   \centerline{\bf Table VIII}

   \vskip 5mm

   \begin{center}

   \begin{tabular}{|c|c|c|c|}\hline
$  T/Tc  $&$     M $&           $<r>$&$   \varepsilon$\\
 \hline
 \multicolumn{4}{|c|}{$M_0=1.044$~~~~~$\gamma=0.2$}\\
  1.007& 3.39&  3.38 &-0.041\\
  1.030 &3.15 & 7.08 &-0.006  \\
  1.065& 3.03 & 20.74 &-0.0003\\   \hline
 \multicolumn{4}{|c|}{$M_0=1.044$~~~~~$ \gamma=1.044$}\\
  1.007 & 3.42&  6.96& -0.005\\\hline

  \end{tabular}

  \end{center}

     \newpage

     \centerline{\bf Table IX}
  \small
    \begin{center}
     \begin{tabular}{|c|c|c|c|}\hline
$T/T_c$&$M $& $  <R> $& $\varepsilon$\\\hline
  \multicolumn{4}{|c|}{$ (s\bar s)_1;~~ M_0=0.69, \gamma=0.20
  $}\\
  1.00 & 1.74&   9.76& -0.0056
  \\
  1.05 & 1.69 &  12.33 & -0.0015
  \\\hline

  \multicolumn{4}{|c|}{$(c\bar c)_3;~~ M_0=0.69, \gamma=0.20
  $
  } \\
  1.00 & 3.49&   5.571 &-0.014
  \\
  1.05  &3.46 &  6.161 &-0.010
  \\
  1.10  &3.42 &  6.931 &-0.008
  \\
  1.15  &3.39 &7.96&-0.005
  \\ \hline

   \multicolumn{4}{|c|}{$(c\bar c)_3;~~ M_0=0.69, \gamma=0.69
   $}  \\
  1.00 & 3.50 &  7.40 &-0.006
  \\
  1.05&   3.46 & 8.66 &-0.004
  \\
  1.10 & 3.43&  10.36&-0.002
  \\
  1.15  & 3.39 & 12.44 &-0.001
  \\ \hline

   \multicolumn{4}{|c|}{$(cg)_3;~~ M_0=0.69, \gamma=0.20
   $} \\
 1.00  &  3.03    &     3.55&-0.071
 \\
 1.05   & 2.97      &   3.81 & -0.057
 \\
 1.10   & 2.92      &   4.13&-0.045
 \\
 1.15   & 2.87      &   4.56&-0.035
 \\ \hline

   \multicolumn{4}{|c|}{$ (cg)_3;~~ M_0=0.69, \gamma=0.69
   $}\\
 1.00 &3.06   & 4.35 &-0.039
 \\
 1.05  &3.00  & 4.80 &-0.029
 \\
 1.10  &2.95  & 5.45 &-0.021
 \\
 1.15  &2.89  & 6.40 &-0.014
 \\ \hline

   \multicolumn{4}{|c|}{$ (gg)_1;~~ M_0=1.0,  \gamma=0.20$}\\
  1.00 & 2.16&     3.72 & -0.042\\
  1.05 & 2.11 &    4.29   &-0.029\\
  1.10&  2.06 &    5.12 &  -0.019
  \\
  1.15 & 2.01  &   6.41 &  -0.011
  \\ \hline

   \multicolumn{4}{|c|}{$ (gg)_1;~~ M_0=0.69, \gamma=0.20
   $}\\
  1.00 & 2.26 &  2.10  &-0.335\\
  1.05 & 2.22&     2.21&-0.289
  \\
  1.10 & 2.17 &    2.34&-0.246
  \\
  1.15 & 2.13 &    2.49&-0.207
  \\ \hline

  \multicolumn{4}{|c|}{$ (gg)_8;~~ M_0=0.69, \gamma=0.20
  $}\\
  1.00& 2.50&   11.27& -0.003
  \\
  1.05& 2.42 &  13.48 &-0.0005
\\ \hline

\end{tabular}

\end{center}

\normalsize

\vfill\eject

\noindent
\begin{center}
{\bf FIGURE CAPTIONS}
\end{center}
\vskip 0.5 cm
\begin{itemize}
\item [\bf Fig.~1]
The behaviour obtained for the static potential $V_1(R,T)$, given
by Eqs. (\ref{V1tot})--(\ref{V1np}), as a function of the distance
$R$, for different values of the temperature $T>T_c$, using for
$B(T)$ and $M$ the central values reported in Table I and for the
parameter $b$ the perturbative estimate (\ref{b-pert}), with the
{\it bare} lattice coupling constant $\alpha_s = 3/(2\pi\beta) \simeq 0.08$,
as explained in section 4.
On the right--hand side the scale is multiplied by 7.5, which corresponds
to $\alpha_s \simeq 0.6$, the characteristic value at large distances,
as explained in section 5.

\item [\bf Fig.~2]
The three-quark potential $V_1(Q\bar Q)(r,T)$ in different color
states for quarks placed in vertices of equilateral triangle with
sides $r$, $vs ~r$ in GeV$^{-1}$.

\item [\bf Fig.~3]
Masses of $c\bar c$  color singlet bound states (in GeV) as
functions of $T/T_c$ for sets of parameters 1-4 from
Eq.(\ref{56}).

\item [\bf Fig.~4]
Binding energies of $c\bar c, L=0$ bound state, defined as
$\varepsilon\equiv M(c\bar c) -2m_c -V_1(\infty, T)$ as functions
of $T/T_c$ for the parameter sets 1-4, Eq. (\ref{56}).

\item [\bf Fig.~5]
Bottomonium $L=0$ masses (in GeV) $vs~ T/T_c$ for parameter sets
2,4, Eq. (\ref{56}), $m_b=4.8$ GeV.

\item [\bf Fig.~6]
Masses of bound states (in GeV) for the systems labelled in Table
VI and for the  potential $V_1(I)~~ vs~~ T/T_c$  with  the
parameter set 4 from Eq. (\ref{56}).
\end{itemize}

\vfill\eject

\pagestyle{empty}

\vfill\eject

\pagestyle{empty}


\begin{figure}
\includegraphics[angle=-90,width=8cm]{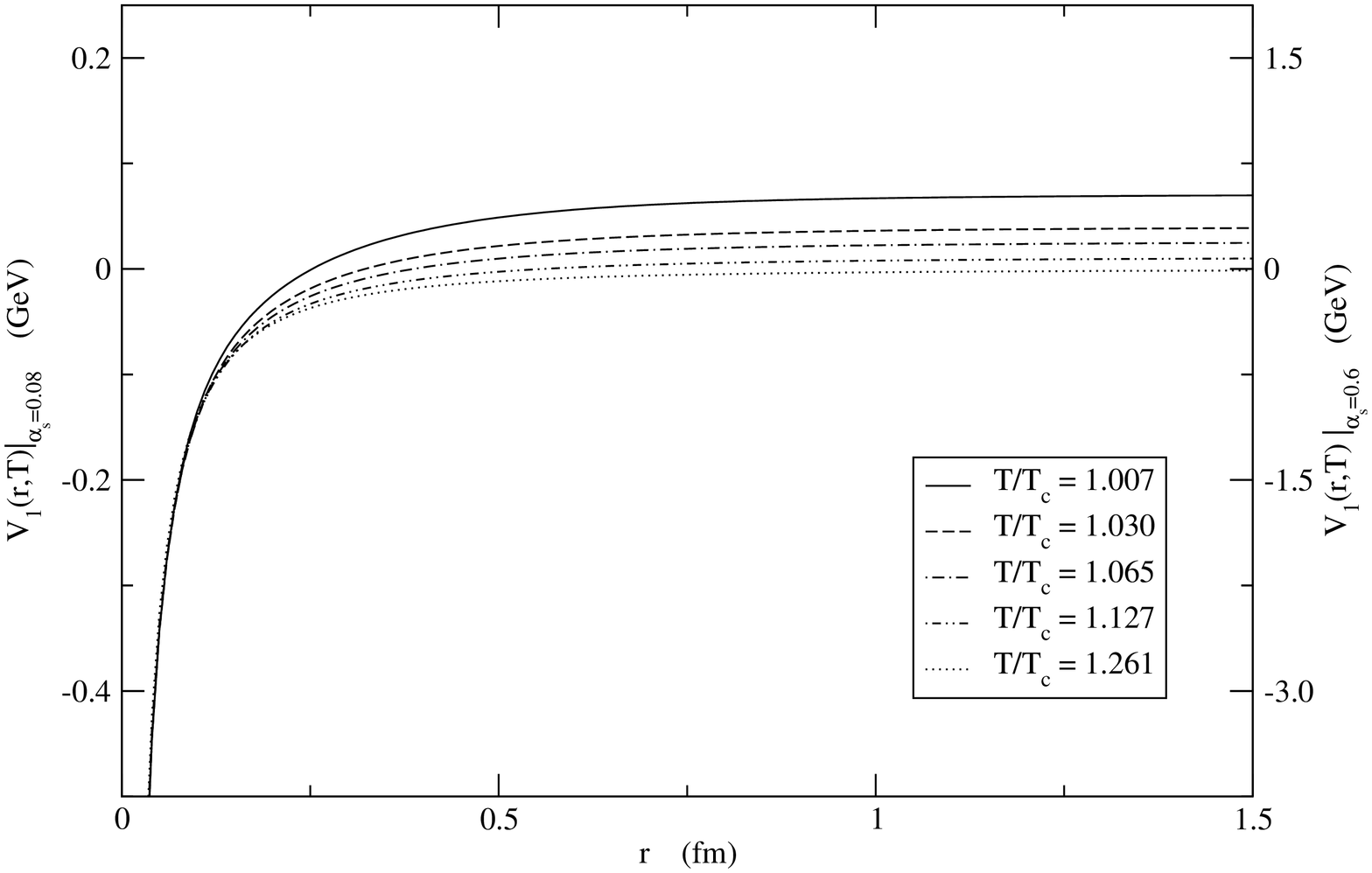}
\caption{}
\end{figure}

\begin{figure}
\includegraphics[width=12cm]{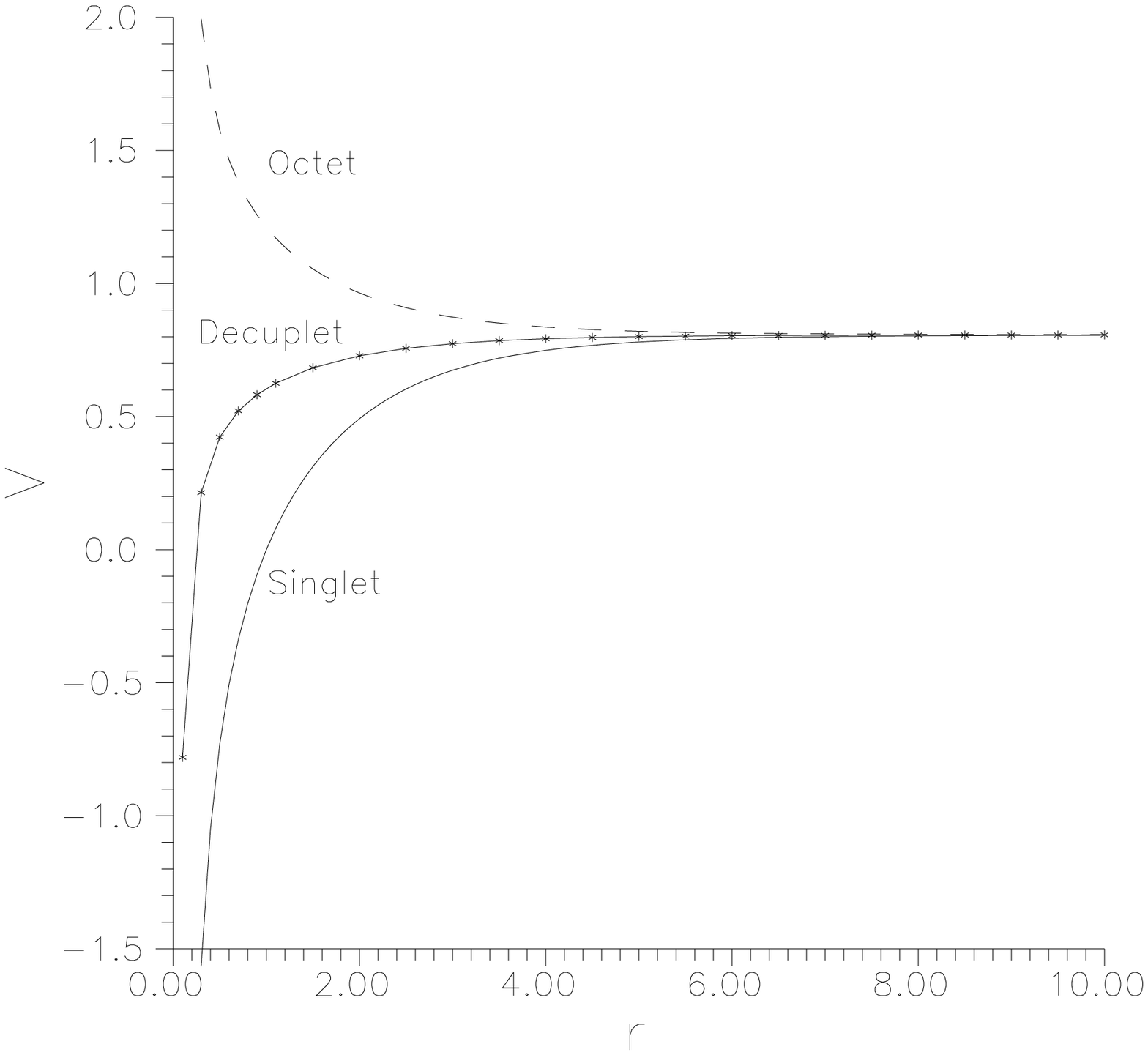}
\caption{}
\end{figure}

\begin{figure}
\includegraphics[width=12cm]{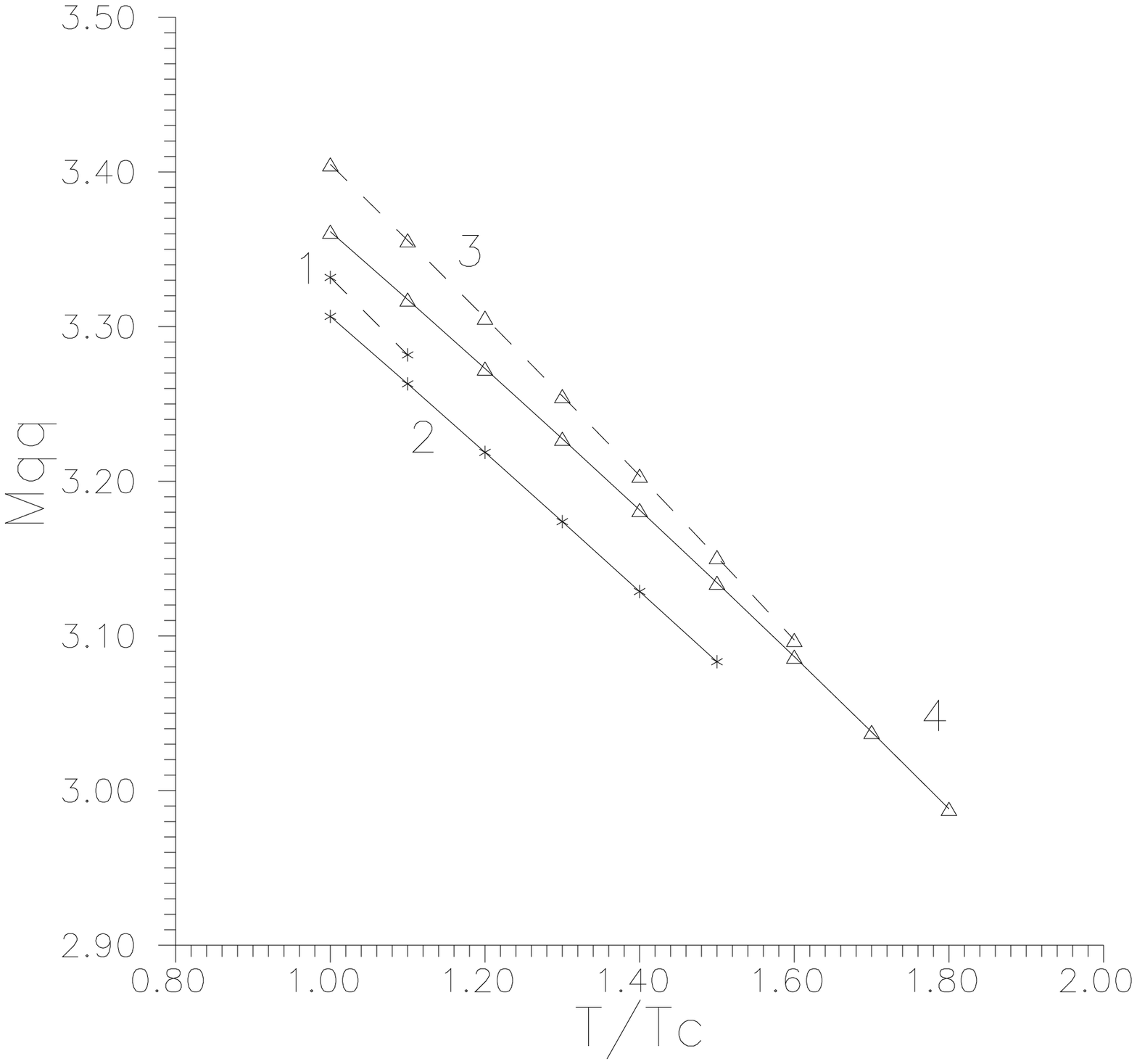}
\caption{}
\end{figure}

\begin{figure}
\includegraphics[width=12cm]{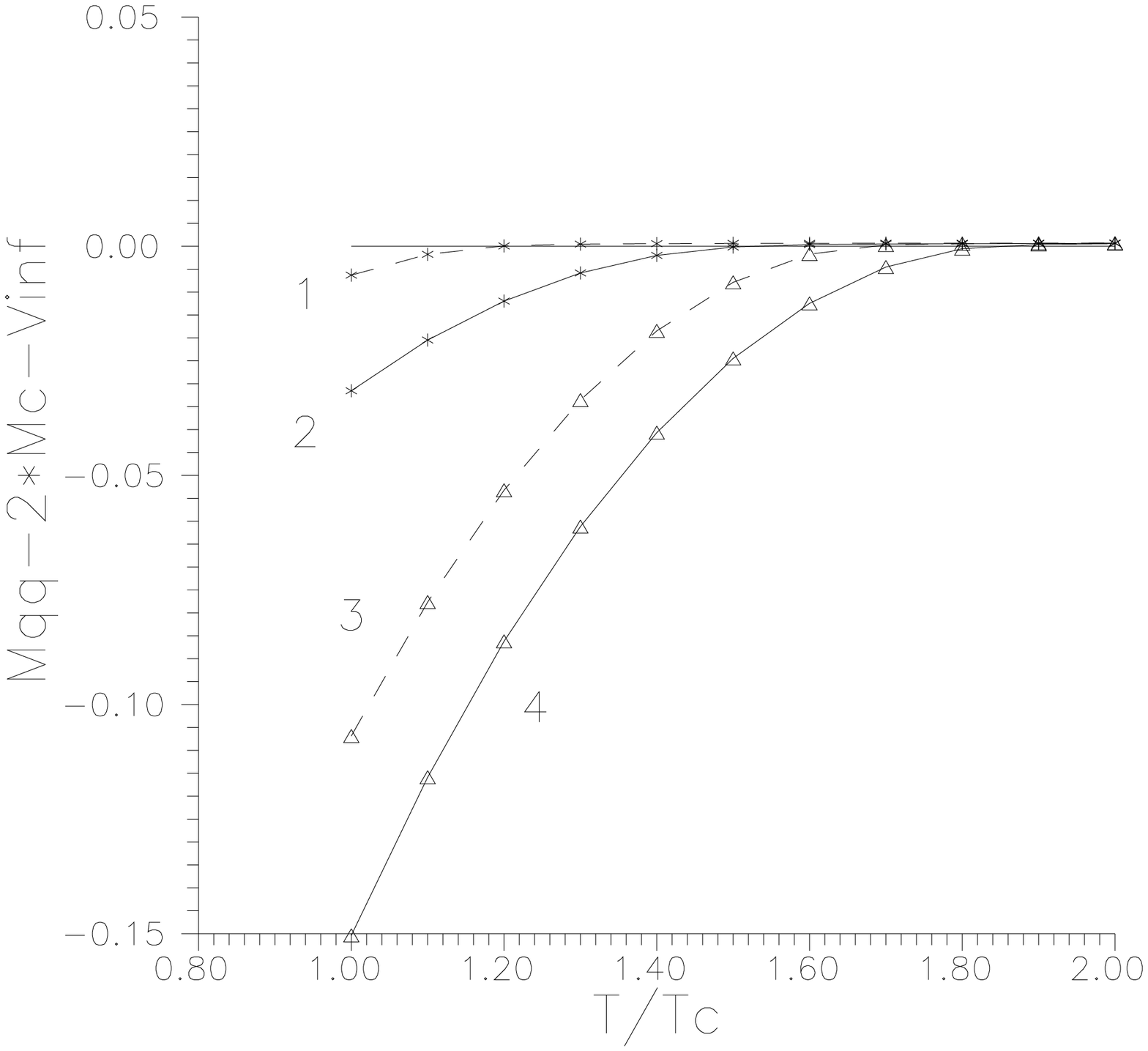}
\caption{}
\end{figure}

\begin{figure}
\includegraphics[width=12cm]{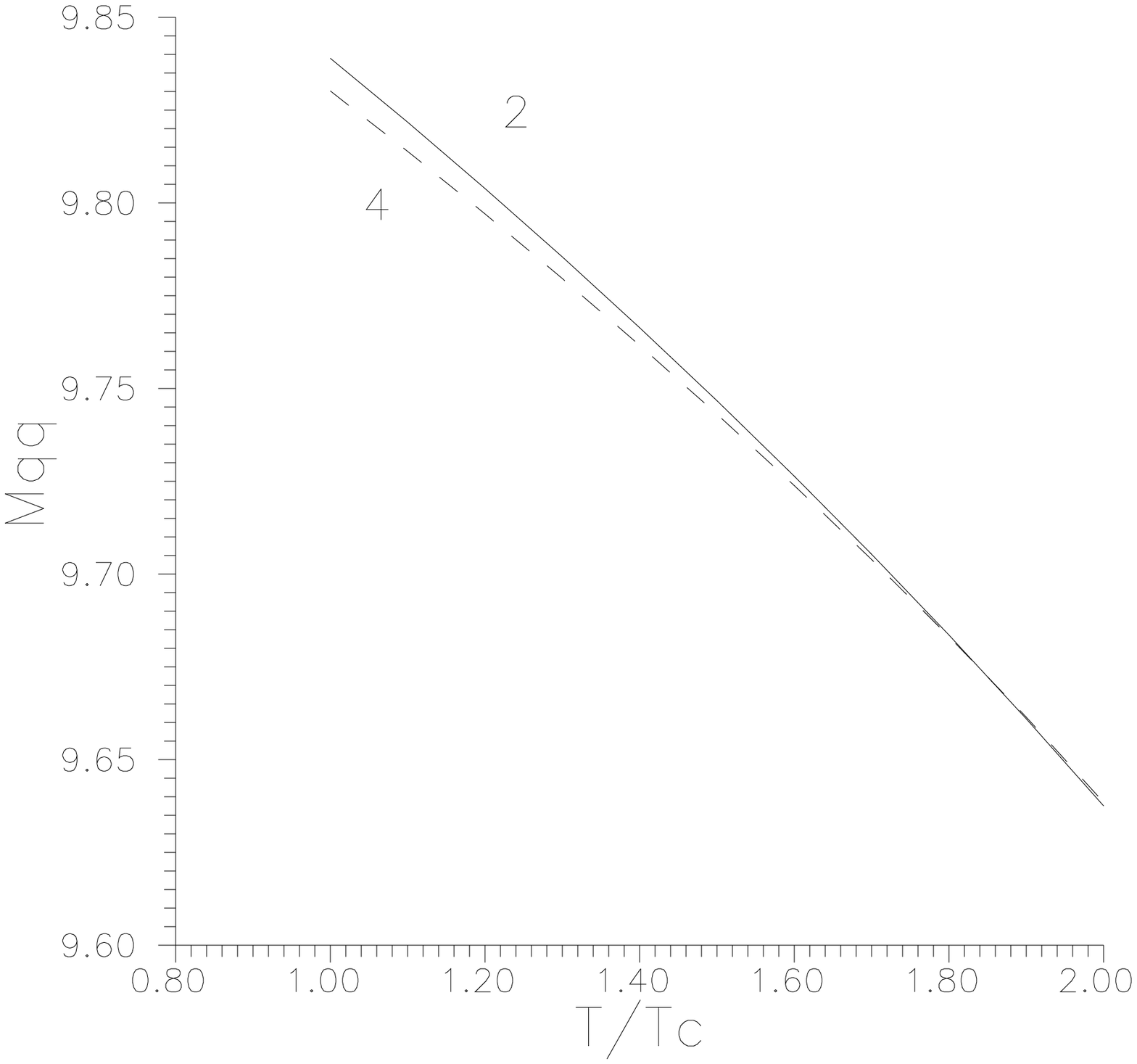}
\caption{}
\end{figure}

\begin{figure}
\includegraphics[width=12cm]{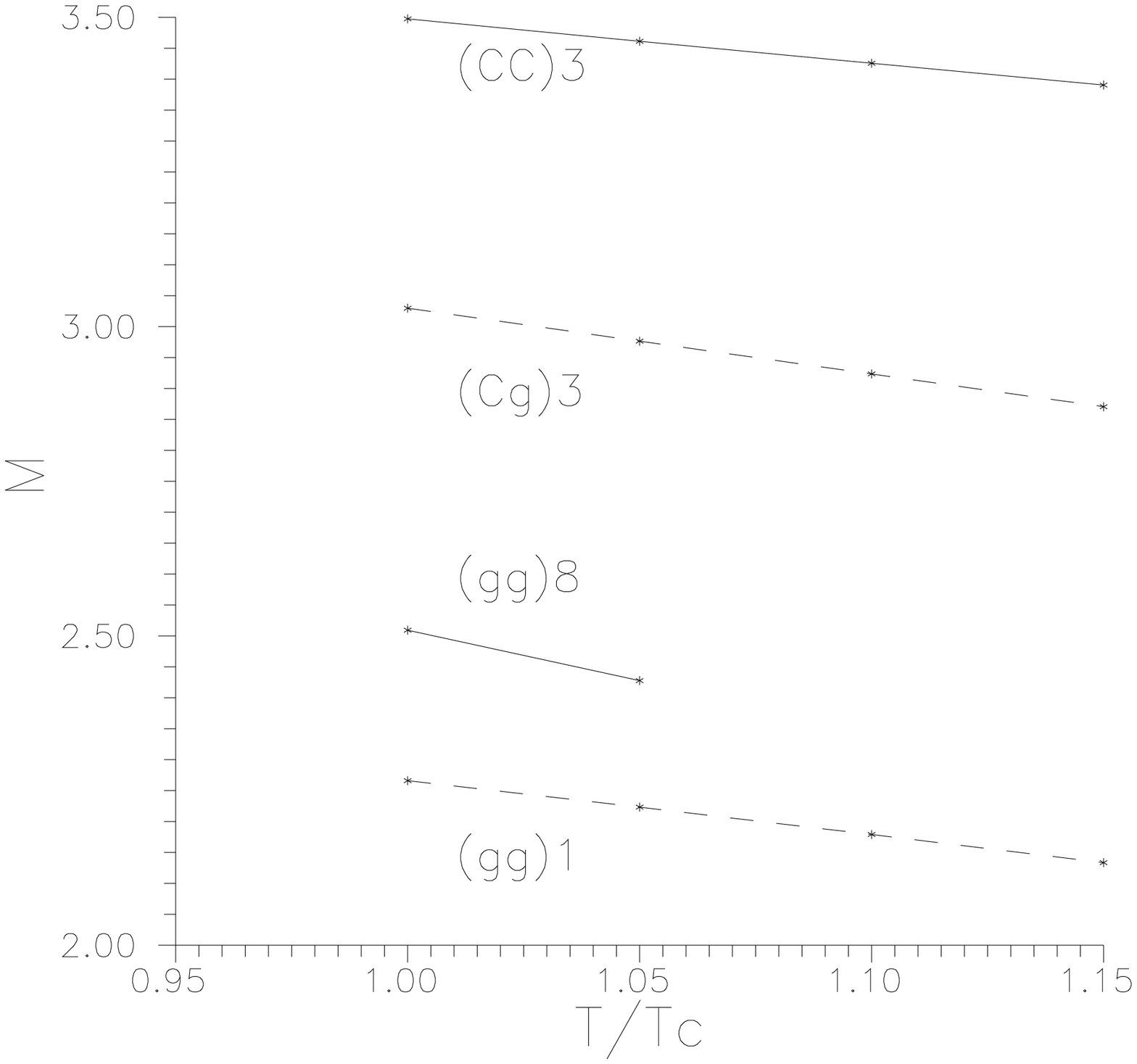}
\caption{}
\end{figure}

\end{document}